\documentclass[fleqn,usenatbib]{mnras}
\usepackage{newtxtext,newtxmath}
\usepackage[T1]{fontenc}
\usepackage{ae,aecompl}
\usepackage{relsize}

\usepackage{graphicx}	
\usepackage{amsmath}	
\usepackage{upgreek}
\usepackage{bm}

\graphicspath{{./}{figures/}}
\newcommand{\Disco}{{\texttt{Disco}}}
\newcommand{\tensorSymbol}[1]{{\overset{\leftrightarrow}{#1}}}

\title[The Terminal Spins of Accreting Objects]{On the Terminal Spins of Accreting Stars and Planets: Boundary Layers}

\author[A. J. Dittmann]{
Alexander J. Dittmann$^{1}$\thanks{E-mail: \href{mailto:dittman@astro.umd.edu}{dittmann@astro.umd.edu}}
\\
$^{1}${Department of Astronomy, University of Maryland, College Park, MD 20742-2421}\\
}

\date{Accepted XXX. Received YYY; in original form ZZZ}

\pubyear{2021}

\begin{document}
\label{firstpage}
\pagerange{\pageref{firstpage}--\pageref{lastpage}}
\maketitle

\begin{abstract}
The origin of the spins of giant planets is an open question in astrophysics. As planets and stars accrete from discs, if the specific angular momentum accreted corresponds to that of a Keplerian orbit at the surface of the object, it is possible for planets and stars to be spun up to near-breakup speeds. However, accretion cannot proceed onto planets and stars in the same way that accretion proceeds through the disk. For example, the magneto-rotational instability cannot operate in the region between the nearly-Keplerian disk and more slowly-rotating surface because of the sign of the angular velocity gradient. Through this boundary layer where the angular velocity sharply changes, mass and angular momentum transport is thought to be driven by acoustic waves generated by global supersonic shear instabilities and vortices. We present the first study of this mechanism for angular momentum transport around rotating stars and planets using 2D vertically-integrated moving-mesh simulations of ideal hydrodynamics. We find that above rotation rates of $\sim0.4-0.6$ times the Keplerian rate at the surface, depending on the gas sound speed, the rate at which angular momentum is transported inwards through the boundary layer by waves decreases by $\sim1-3$ orders of magnitude. We also find that the accretion rate through the boundary layer decreases commensurately and becomes less variable for faster-rotating objects. 
Our results provide a purely hydrodynamic mechanism for limiting the spins of accreting planets and stars to factors of a few less than the breakup speed. 
\end{abstract}

\begin{keywords}
accretion, accretion discs -- hydrodynamics -- planet formation
\end{keywords}


\section{Introduction}
The terminal spins of accreting objects are of great interest in both planetary science and high-energy astrophysics. For example, as massive planetary cores accrete from protoplanetary discs, they are expected to spin up to near-critical rates \citep[e.g.][]{2008ApJ...685.1220M}. However, planets in our own Solar System spin at considerably less than the critical value (e.g. $\Omega_*/\Omega_{\rm crit}\sim\{0.3,~0.4\}$ for Jupiter and Saturn respectively), where $\Omega_{\rm crit}$ is defined as the Keplerian angular frequency at the surface 
and $\Omega_*$ is the angular velocity at the surface. The observed sub-critical rotation rates could be attributed to effects such as magnetic braking in the presence of strong planetary magnetic fields \citep[e.g.][]{2011AJ....141...51L,2018AJ....155..178B,2020MNRAS.491L..34G}. The observed present-day magnetic field strength of Jupiter is orders of magnitude too low for magnetic braking to be significant \citep{1974JGR....79.3501S}, although proto-Jupiters may have stronger magnetic fields \citep{2009Natur.457..167C}. Limited spin measurements of extrasolar planets suggest similar ratios of $\Omega_*/\Omega_{\rm crit}\sim0.08-0.3$ \citep{2018NatAs...2..138B}. 
The fastest-spinning known pulsar, thought to have been spun-up through accretion, also rotates at $\Omega_*/\Omega_{\rm crit}\sim0.43$ assuming a mass of $1.4~M_\odot$ and radius of 12 kilometres \citep{2006Sci...311.1901H}. 
Additionally, stars may accrete rapidly after forming in or being captured into accretion discs around active galactic nuclei (AGNs) \citep[e.g][]{1980SvAL....6..357K,2003ApJ...590L..33L,2004ApJ...608..108G,2020MNRAS.493.3732D}, often reaching near-critical spins and masses of $\sim10~M_\odot$ near the ends of their lives \citep{2021ApJ...910...94C,2021ApJ...914..105J,2021arXiv210212484D}. If such spins are realised, long gamma ray bursts are expected in AGN discs, although their appearance would strongly depend on local disc properties \citep[e.g.][]{2021ApJ...906L...7P,2021ApJ...911L..19Z}.

However, for objects such as stars and planets, which unlike black holes possess surfaces, the physical mechanisms for the transport of angular momentum from an accretion disk into the object are not straightforward. Specifically, the angular velocity of material near the surface of the object is much smaller than the Keplerian angular velocity at that radius. Material must therefore accrete through a boundary layer
where angular velocity increases with radius $(d\Omega/dr>0)$. However, the magnetorotational instability \citep{velikhov59,1960PNAS...46..253C}, which is thought to drive accretion in many astrophysical disks \citep{1991ApJ...376..214B}, can only operate when $d\Omega/dr<0$. Thus, a different mechanism for angular momentum transport is necessary within the boundary layer. 

Acoustic waves, generated by instabilities due to supersonic shear in the boundary layer, are one potential mechanism for angular momentum transport through the boundary layer \citep{2012ApJ...752..115B,2012ApJ...760...22B,2013ApJ...770...67B,2013ApJ...770...68B}. In this case, acoustic waves are excited in the boundary layer and propagate into both the disk and central object, transporting angular momentum globally. Notably, because this angular momentum transport is nonlocal, it cannot be emulated by local viscous models where angular momentum transport occurs through shear stresses as assumed in some models of accretion disks and boundary layers \citep[e.g.][]{1973A&A....24..337S,1995ApJ...442..337P,1999ApJ...521..650B}. However, in some cases where acoustic waves are not sufficient to transport through the boundary layer all of the angular momentum accreted through the disk, `belts' of angular momentum may form without appreciably spinning up the central object \citep{2018MNRAS.479.1528B}. Additionally, previous work on acoustic waves generated within boundary layers has focused on the case of non-spinning central objects \citep{2012ApJ...752..115B,2012ApJ...760...22B,2013ApJ...770...67B,2013ApJ...770...68B,2018MNRAS.479.1528B,2021arXiv210312119C}, and is thus not necessarily applicable to objects possessing non-negligible spins. We note that other recent work has examined the terminal spins of accreting planets \citep{2020arXiv201206641D}, finding maximum angular velocities of $\Omega/\Omega_{\rm crit}\sim0.6-0.8$, albeit using 2D azimuthally-symmetric $(r-\theta)$ simulations with a constant$-\nu$ viscosity prescription which could not accurately capture angular momentum transport through the boundary layer due to acoustic waves. 

Herein we aim to study angular momentum transport through boundary layers around rotating and nonrotating objects, and determine how the efficacy of acoustic modes at spinning up objects depends on the rotation rate of the central object using simulations of vertically-integrated inviscid hydrodynamics. Although the central objects in such systems could be planets, neutron stars, white dwarfs, protoplanets, stars, or protostars, we will typically refer to them simply as ``stars." We review some previous results on boundary layers in Section \ref{sec:boundaryBasics}, and present our numerical methods and initial conditions in Section \ref{sec:methods}. We present the results of our simulations in Section \ref{sec:results}, discuss our results in Section \ref{sec:discuss}, and conclude in Section \ref{sec:conclusions}.

\section{Acoustic modes}\label{sec:boundaryBasics}
We begin by reviewing some of the basics of acoustic modes in boundary layers, which have been discussed previously in detail \citep{2012ApJ...752..115B,2012ApJ...760...22B,2013ApJ...770...67B}. In disk-star boundary layers, there is typically a large velocity difference between approximately Keplerian material in the disk and more slowly-rotating material in the star. For nearly incompressible gas with velocities typically much less than the sound speed, such shear can lead to the development of the Kelvin-Helmholtz instability. However, the Kelvin-Helmholtz instability is suppressed at high Mach numbers \citep{1958JFM.....4..538M}, where instead a global sonic instability occurs \citep{1988MNRAS.231..795G,2012ApJ...752..115B}. \citet{2012ApJ...752..115B} also showed that the growth rate of the sonic instability is independent of the density contrast across the shear layer. 

Various modes within such supersonic shear layers can have corotation resonances where their radial wavenumber becomes zero. These acoustic modes have globally conserved actions, or pseudo-energies, which are positive when a mode travels with a pattern speed larger than the fluid velocity, but negative when modes travel with speed less than the fluid velocity, and thus the action changes sign at corotation \citep{1987MNRAS.228....1N}. Acoustic waves can partially reflect off of corotation resonances, with some action tunnelling through, in which case global conservation of action leads to amplification and then instability of various modes when trapped between multiple corotation resonances or between a corotation resonance and reflecting boundary \citep{1987MNRAS.228....1N,1988MNRAS.231..795G,2012ApJ...752..115B}. In this spirit, we will visualise the spatial morphology of various modes through the quantity $rv_r\sqrt{\Sigma}$ as a proxy for acoustic wave action, which is a useful quantity because both the energy and angular momentum currents of acoustic waves are conserved and of order $r^2\Sigma\delta \mathbf{v}^2$, where $r$ is a radius, $\Sigma$ the surface density, $v_r$ is the radial component of the fluid velocity, and $\delta \mathbf{v}$ is a velocity perturbation associated with a wave. 

In the terminology of \citet{2012ApJ...752..115B}, we typically observe `upper' and `lower' modes in our simulations, referring to branches of their derived dispersion relations. The lower mode visibly dominates a large number of our simulations, and is characterised by nearly plane-parallel propagation of acoustic waves within the star such that the radial and azimuthal wavenumbers satisfy $k_\phi\gg k_r$, but non-negligible $k_r$ in the disk. One particularly striking example is the snapshot from our $\Omega_0=0.4,~\mathcal{M}=8$ simulation after 300 Keplerian orbital periods at the stellar surface shown in Figure \ref{fig:actionGrid}, where $\Omega_0\equiv\Omega_*/\Omega_K(r_*)$, the ratio of the stellar angular velocity to the Keplerian angular velocity at the stellar surface and $\mathcal{M}$ is the ratio between the Keplerian linear velocity and sound speed at $r=r_*$. In additional to the near-radial `stripes' of $rv_r\sqrt{\Sigma}$ within the star indicative of a lower mode, the effects of corotation resonances both in the disk and boundary layer are visible, as $v_r$ changes sign at the corotation resonance near the disk-boundary layer transition, and again at the corotation resonance in the disk. Intersection between outward-propagating waves from the boundary layer and inward-propagating waves from the disk corotation resonance result in a clear crisscrossed pattern, and the wave amplitude between the two resonances is clearly larger than that of the waves which reach the outer disk.

\begin{figure*}
\includegraphics[width=\linewidth]{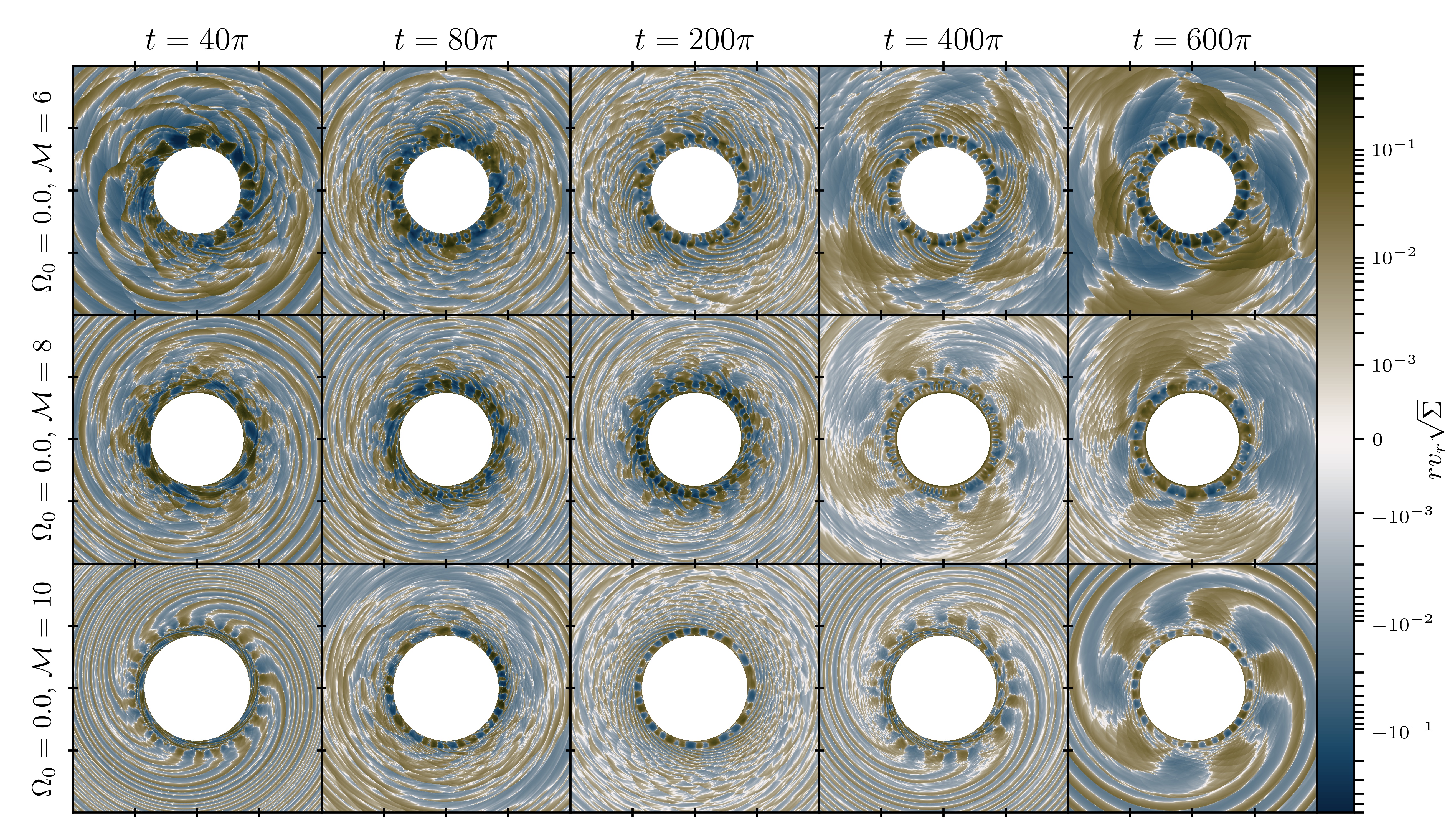}
\caption{Snapshots of $rv_r\sqrt{\Sigma}$ over time for simulations with $\Omega_0=0$. The first, second, and third rows show plots from the $\mathcal{M}=\{6,8,10\}$ simulations respectively. The first through fifth columns are grouped by simulation time, at $t=\{40,80,200,400,600\}\pi$.
}
\label{fig:actionTimeGrid}
\end{figure*}

Upper modes typically dominate our simulations only at early times, and are characterised by propagation with $k_\phi\gg k_r$ in the disk near the boundary layer but non-negligible $k_r$ within the star. However, further into the disk from the boundary layer, these modes are wound by differential rotation and acquire significant $k_r$ in the disk as well. This structure is visible in the snapshot of our $\Omega_0=0,~\mathcal{M}=10$ simulation after 20 Keplerian orbital periods at the stellar surface displayed in the bottom-left panel of Figure \ref{fig:actionTimeGrid}.

In addition to acoustic modes seeded by the sonic instability, \citet{2021arXiv210312119C} identified a number of acoustic modes driven by vortices in the boundary layer. \citet{2021arXiv210312119C} found that global waves in the disk were typically driven by localised vorticies in the boundary layer which extend over small ranges in azimuth, as well as more extended `roll' structures which cover large swaths in azimuth. To study these structures, we also investigate the vorticity ($\omega=\nabla\times \mathbf{v}$) and vortensity ($\omega/\Sigma$, otherwise known as potential vorticity), as well as the difference in vortensity from that in our initial condition $(\omega_0/\Sigma_0)$ which is a useful comparison due to the conservation of vortensity along streamlines in the absence of shocks and dissipation. We identify modes driven by both isolated and extended vortices, such as our $\Omega_0=0.8,~\mathcal{M}=6$ and $\Omega_0=0,~\mathcal{M}=10$ simulations shown in Figure \ref{fig:vorticityGrid}.

To analyse momentum transport in our simulations in order to assess how the spin-up of stars depends on their rotation rate, we calculate the angular momentum current due to stresses
\begin{equation}
C_s = 2\pi r^2 \Sigma_0 \left( \langle v_r v_\phi \rangle - \langle v_r \rangle \langle v_\phi \rangle \right),
\end{equation}
where $\langle ...\rangle$ represents a mass-weighted average in $\phi$ and $\Sigma_0$ is the $\phi-$averaged surface density at a given radius. We note that $C_s$ is related to the angular momentum flux $F_s=C_s/2\pi r$. We also note that the rate of change of angular momentum through an annulus is the difference between the angular momentum currents at the inner and outer edges of that annulus, and that negative values of $C_s$ indicates that angular momentum is being transported to smaller radii. We focus on $C_s$ rather than the full angular momentum current $C_l = 2\pi r^2 \Sigma_0\langle v_r v_\phi \rangle$ because we find that numerical effects make this difficult to measure accurately within stars with $\Omega_0\neq0$. However, we have found that $C_s$ and $C_l$ follow similar trends with $\Omega_0$ and $\mathcal{M}$ in the disk and boundary layer, so focusing on $C_s$ does not affect our conclusions.

\section{Methods}\label{sec:methods}
This study utilises the moving-mesh code \Disco{} \citep{2016ApJS..226....2D}
to solve the 2D ($r,~\phi$) equations of vertically-integrated inviscid isothermal hydrodynamics:

\begin{align}
\partial_t\Sigma + \bm{\nabla}\!\cdot\!(\Sigma\mathbf{v}) &= 0 \label{eq:continuity} \\
\partial_t(\Sigma\mathbf{v}) + \bm{\nabla}\!\cdot\!(\Sigma\mathbf{v}\mathbf{v}+\Pi\tensorSymbol{I})
&= -\Sigma\bm{\nabla}\Phi \label{eq:momentum} \\
\Pi&=c_s^2\Sigma,
\end{align}
where $\Pi$ is the vertically-integrated pressure, $\Sigma$ is the surface density, $\mathbf{v}$ is the fluid velocity vector, $\tensorSymbol{I}$ is the identity tensor, $c_s$ is the isothermal sound speed, and $\Phi$ is a gravitational potential. We assume that the accretion disk has negligible mass compared to the star, and thus neglect self-gravity. We further assume a gravitational potential
\begin{equation}
\Phi=-\frac{1}{r}
\end{equation}
which is appropriate near and beyond the surfaces of centrally-concentrated bodies such as stars and planets. We hold the sound speed fixed globally to
\begin{equation}
c_s = \frac{\sqrt{-\Phi(r_*)}}{\mathcal{M}},
\end{equation}
where $\mathcal{M}$ is a characteristic Mach number, taken to be $\mathcal{M}=\{6,8,10\}$ in this work, and $r_*$ is the radius of the stellar surface. 
The Mach numbers considered here are appropriate for some circumstellar and circumplanetary disks for objects embedded in global accretion disks, although the full range of possible Mach numbers could vary by orders of magnitude depending on global accretion disk aspect ratios and the mass ratio between the central and embedded objects for a given accretion disk. Our results do not show systematic trends as a function of $\mathcal{M}$, so they likely apply to a wider range of disk properties, although our vertically-averaged numerical treatment is a poor approximation for thick disks with $\mathcal{M}\sim1$, so our results should be extended to lower Mach numbers with caution.

Our 2D domain is decomposed into standard cylindrical $(r,~\phi)$ polar coordinates. The radial grid was spaced logarithmically from $r_{\rm in}$ to $r_{\rm out}=3$, using nondimensional units where the radius of the central object is $r_*=1,$ and the Keplerian angular velocity at the surface $\Omega_K(r_*)=1$. Because we ignore the self-gravity of the fluid, we are free to arbitrarily rescale the surface density without loss of generality, and therefore chose units where $\Sigma_0=\Sigma(r_{\rm out})=1$. We summarise our choice of units in Table \ref{tab:units}. The inner radius $r_{\rm in}$ is varied depending on $\mathcal{M}$, with $r_{\rm in}=\{0.7,0.75,0.85\}$ for $\mathcal{M}=\{6,8,10\}$ respectively. Each annulus of our grid rotated with the average angular velocity of the fluid within that annulus. Our fiducial grid resolution was $N_r=2048$ radial zones, with the number of zones in $\phi$ chosen to maintain a cell aspect ratio as close to $\sim1$ as possible, ranging from $N_\phi\sim8860$ to $N_\phi\sim10200$ depending on $r_{\rm in}$. 

\begin{table}
\begin{centering}
\begin{tabular}{cc}
Physical quantity & unit \\ \hline
length & $R_*$ \\
time & $\sqrt{R_*^3/GM_*}=\Omega_*^{-1}$ \\ 
surface density & $\Sigma_0$ \\
\end{tabular}
\caption{The base units used when reporting nondimensional values, where $R_*$ is the radius of the surface of the accreting object, $\Omega_*$ is the Keplerian angular velocity at $R_*$, and $\Sigma_0$ is the initial surface density of the accretion disk. For example, the units of angular momentum are $\Sigma_0R_*^4\Omega_*$.}
\end{centering}
\label{tab:units}
\end{table}

\begin{figure}
\includegraphics[width=\columnwidth]{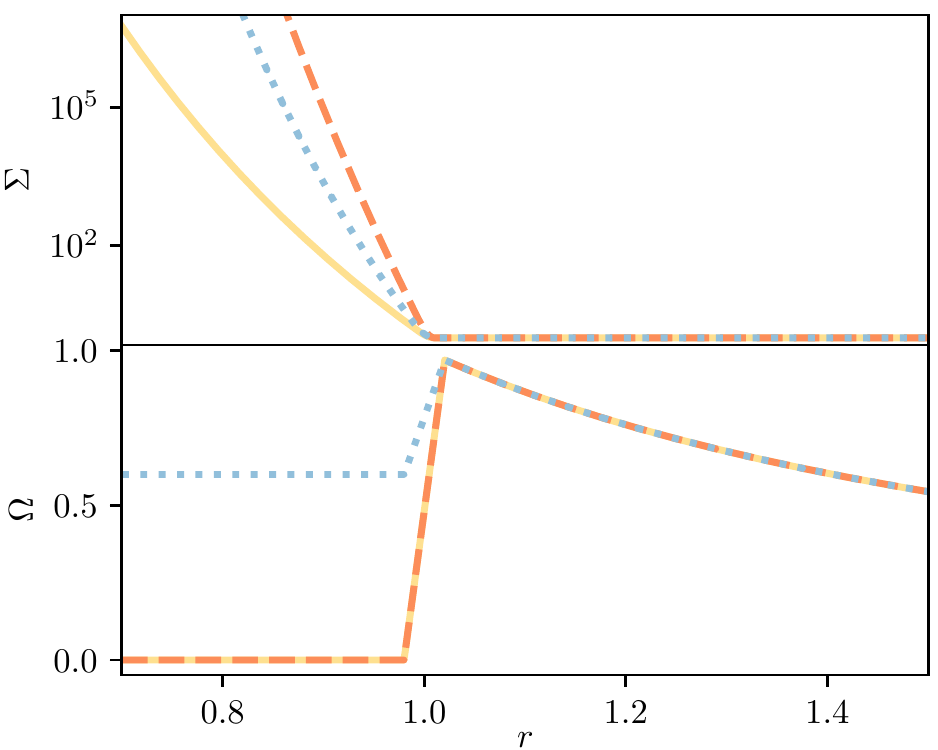}
\caption{Initial conditions for three simulations. The top panel shows surface density, while the bottom panel shows the angular velocity. Yellow solid lines plot the initial conditions for the $\mathcal{M}=6,~\Omega_0=0$ simulation, orange dashed lines show initial conditions for the $\mathcal{M}=10,~\Omega_0=0$ simulation, and blue dotted lines show the initial conditions for the $\mathcal{M}=10,~\Omega_0=0.6$ simulation.}
\label{fig:init}
\end{figure}

We initialise our simulations by specifying an axisymmetric profile for the angular velocity $\Omega(r)$ which is Keplerian outside of the boundary layer and rigidly rotating $(\Omega(r)=\Omega_0)$ within the central object, with a small region linearly interpolating between the two constituting the initial boundary layer:
\begin{equation}
\Omega(r)=
\left\{
\begin{array}{ll}
      r^{-3/2}    & r > r_*+\delta r\\
      \Omega_0 + \left[\left(r_*+\delta r\right)^{-3/2}-\Omega_0\right]\frac{r+\delta_r-r_*}{2\delta r} & r_*-\delta r \leq r \leq r + \delta r\\
      \Omega_0 & r < r_*-\delta r. \\
\end{array}
\right
.
\end{equation}
We fix $\delta r=0.01$ throughout this work. We then specify the surface density profile such that the initial condition is in hydrostatic equilibrium, satisfying
\begin{equation}
\frac{1}{\Sigma}\frac{d\Pi}{dr}=-\frac{d\Phi}{dr}+\Omega^2r,
\end{equation}
which results in constant surface density throughout the disk and $\Sigma\propto\exp{(\mathcal{M}^2/r)}$ within the star for $\Omega_0=0,$ and less-steep surface density profiles for larger $\Omega_0$. 
We resolve length scale over which density changes $(\sim1/\mathcal{M}^2)$ using $\sim\{39,23,16\}$ cells at $r=r_*$ for $\Omega_0=0,~\mathcal{M}=\{6,8,10\}$ respectively. 
Along with varying $\mathcal{M},$ we also vary $\Omega_0=\{0, 0.4, 0.6. 0.8\}$, while previous investigations have focused on the $\Omega_0=0$ case alone \citep[e.g.][]{2012ApJ...760...22B,2013ApJ...770...67B,2013ApJ...770...68B,2021arXiv210312119C}. Sample initial conditions are plotted in Figure \ref{fig:init}. Although our initial conditions are specified analytically through hydrostatic equilibrium, they do not necessarily satisfy \emph{numerical} hydostatic equilibrium \citep[e.g.][]{2002ApJS..143..539Z}, although the simulation quickly settles into equilibrium within a few orbital periods at the stellar surface. Rather than explicitly seeding instability with velocity perturbations as is done when using fixed-grid codes, we use grid-scale noise associated with our moving mesh to seed instabilities in the shear layer. We employ Dirichlet boundary conditions, fixing the values in the outermost cells to their values in the initial condition.

We have performed a number of tests that suggest our fiducial resolution parameters above are sufficient, considering that in simulations of inviscid hydrodynamics all dissipation is purely numerical. For example, we carried out a limited set of simulations at $N_r=3072$ and $N_r=4096$, and found that values of the angular momentum current within the star varied on the order of tens of percent, but not monotonically with resolution. Similarly, at higher values of $\Omega_0$, the surface density contrast between $\Sigma(r_{\rm in})$ and $\Sigma(r_*)$ becomes smaller. We carried out a small set of simulations using a smaller value of $r_{\rm in}$ such that $\Sigma(r_{\rm in})/\Sigma(r)$ was held constant, and found that the angular momentum current within the star also varied by tens of percent, although in different directions relative to the fiducial case at late and early times. In all cases, we observed the same general trends with increasing $\Omega_0$ as in our simulations with fiducial $r_{\rm in}$ and $N_r$. Additionally, the general trends we observe with $\Omega_0$ are much larger than run-to-run variations due to resolution parameters, so our simulations are sufficiently resolved for our current purposes. 

\section{Results}\label{sec:results}

\begin{figure}
\includegraphics[width=\columnwidth]{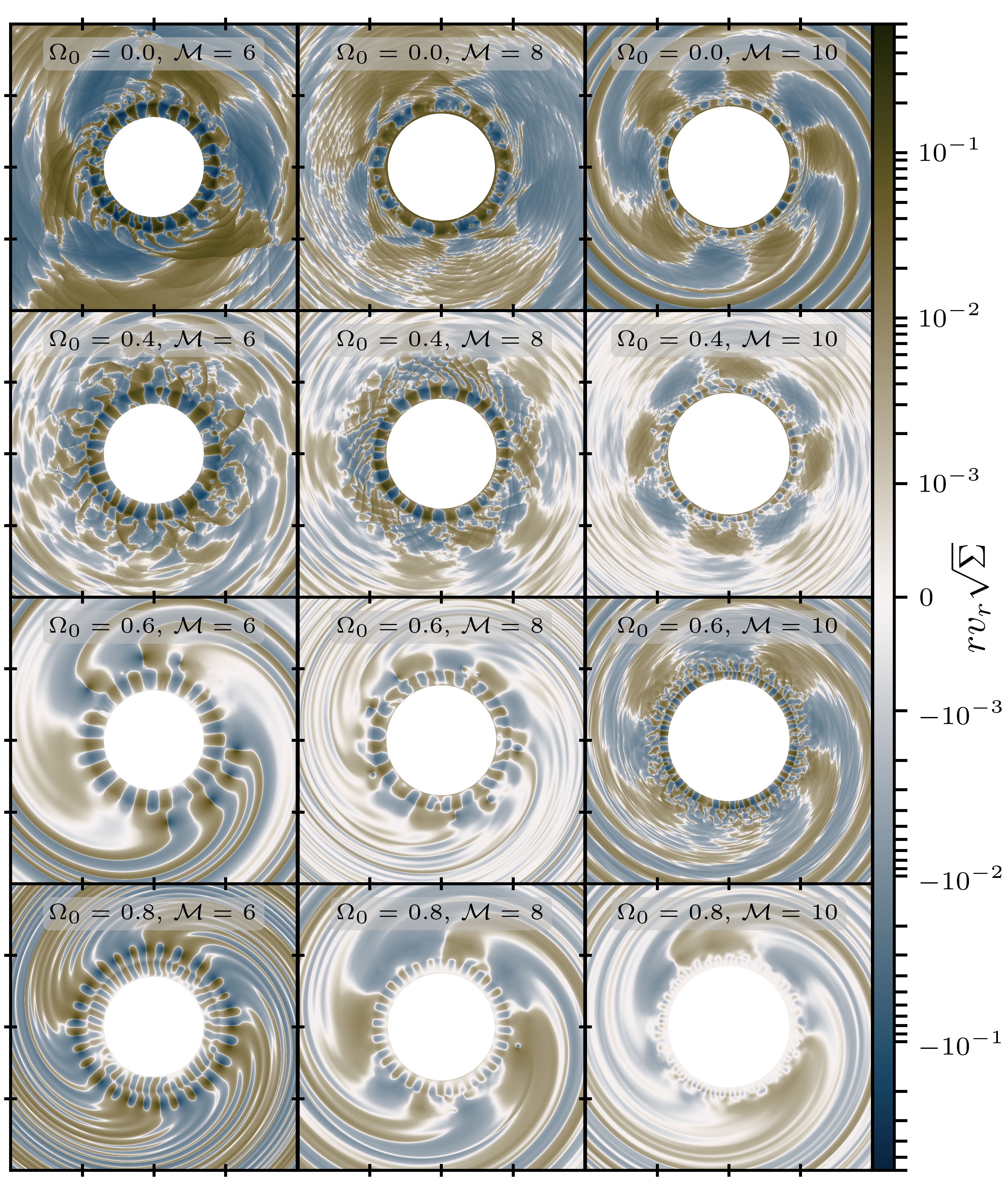}
\caption{Snapshots of $rv_r\sqrt{\Sigma}$ at $t=600\pi$ for each simulation. The columns are grouped by Mach number ($\mathcal{M}=\{6,8,10\}$), while the rows are grouped by stellar rotation rate ($\Omega_0=\{0,0.4,0.6,0.8\}$).}
\label{fig:actionGrid}
\end{figure}

\begin{figure*}
\includegraphics[width=\linewidth]{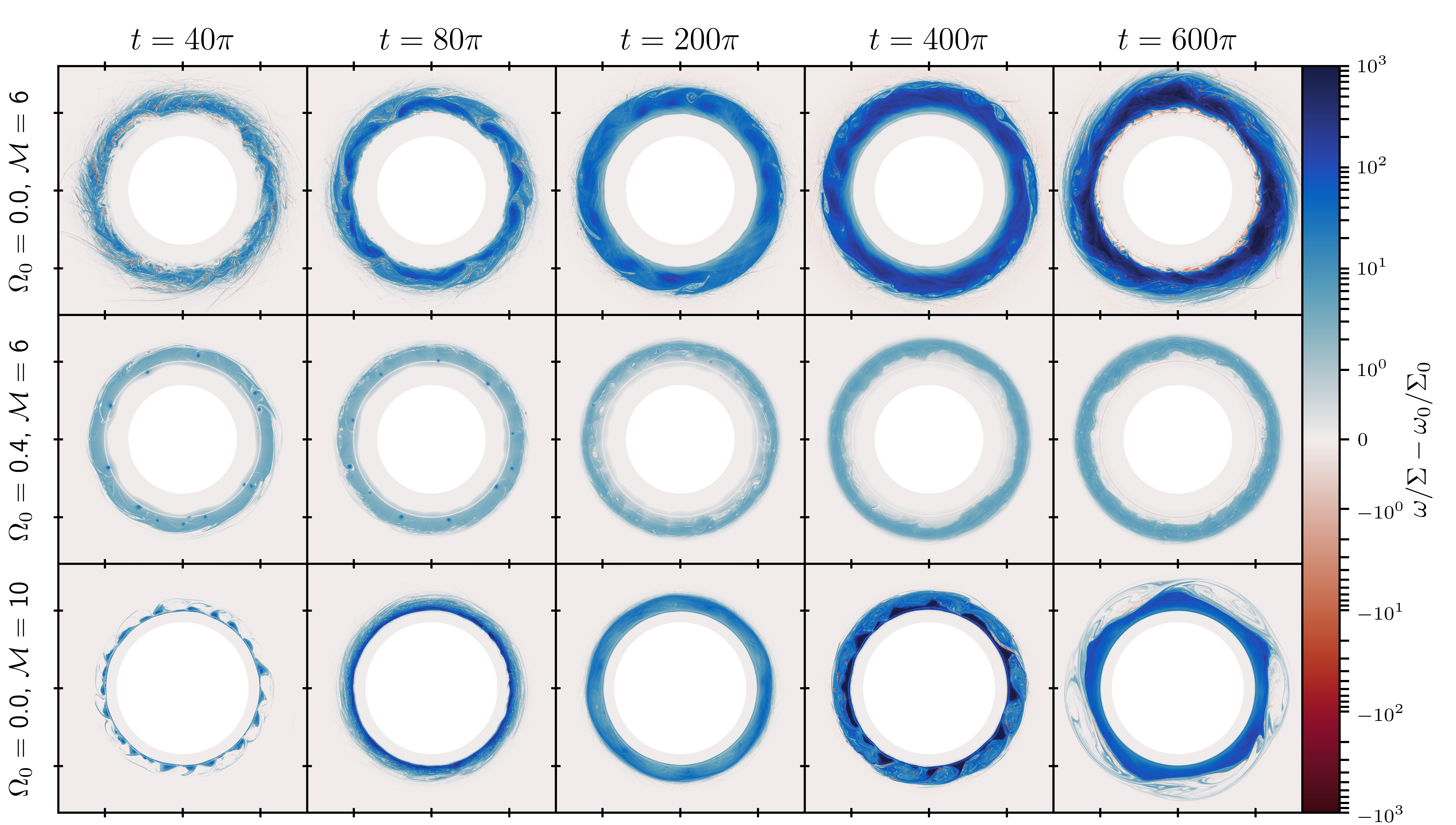}
\caption{Snapshots of the change in vortensity from that of the initial condition $\omega/\Sigma-\omega_0/\Sigma_0$ over time for the $\Omega_0=0,~\mathcal{M}=6$, $\Omega_0=0.4,~\mathcal{M}=6$, and $\Omega_0=0,~\mathcal{M}=10$ simulations. 
The first through fifth columns are grouped by simulation time, at $t=\{40,80,200,400,600\}\pi$. Vortensity tends to move from small to large scales over time, and can be generated by shocks as shown in the first row.}
\label{fig:vortensityTimeGrid}
\end{figure*}

\begin{figure}
\includegraphics[width=0.995\linewidth]{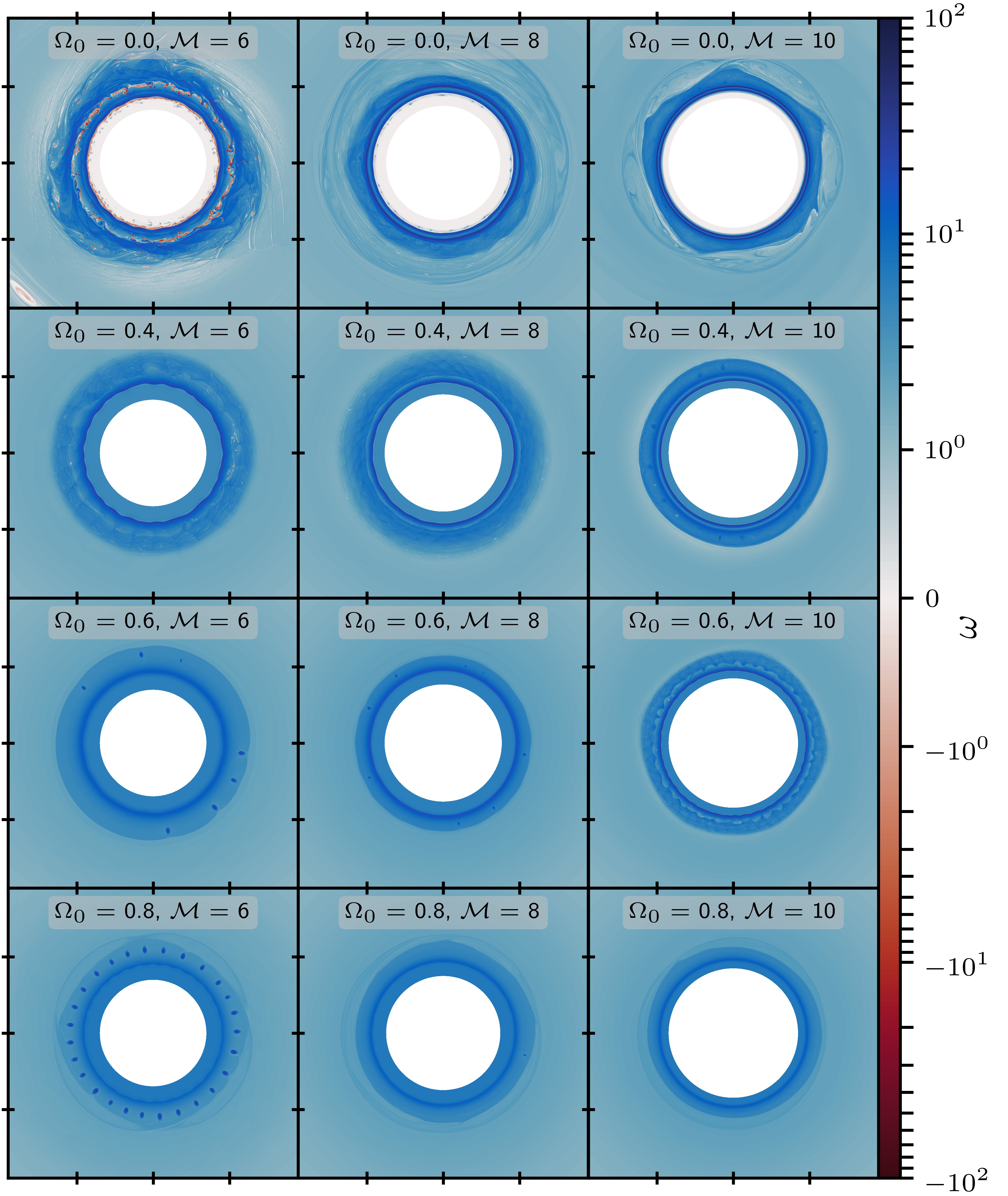}
\caption{Snapshots of vorticity at $t=600\pi$ for each simulation. The columns are grouped by Mach number ($\mathcal{M}=\{6,8,10\}$), while the rows are grouped by stellar rotation rate ($\Omega_0=\{0,0.4,0.6,0.8\}$).}
\label{fig:vorticityGrid}
\end{figure}

We begin by giving a qualitative overview of how the boundary layers in our simulations evolve over time, and how this evolution depends on the rotation rate of the star. Because acoustic modes can be driven by both vortices in the boundary layer and by the sonic instability, we examine the evolution over time of $rv_r\sqrt{\Sigma}$ in Figures \ref{fig:actionTimeGrid} and \ref{fig:actionGrid}, vortensity in Figure \ref{fig:vortensityTimeGrid}, and vorticity in Figure \ref{fig:vorticityGrid} to illustrate their general behaviour. Snapshots of $rv_r\sqrt{\Sigma}$ at various times over the course of our $\Omega_0=0$ simulations are displayed in Figure \ref{fig:actionTimeGrid}. Broadly speaking, higher$-m$ modes tend to dominate at early times, while by $t=600\pi$ lower$-m$ modes almost always dominate. As shown in Figure \ref{fig:actionGrid}, low$-m$ modes also dominate at late times in the $\Omega_0\neq0$ simulations. Overall, more modes are visually apparent in the lower-$\mathcal{M}$ simulations at early times, while at similar times spiral waves are more visible in the disk in the $\mathcal{M}=10$ simulation. Although Figure \ref{fig:actionTimeGrid} is not sufficient to make quantitative statements about the amplitude of acoustic waves and the amount of angular momentum transported, it is clear that the amplitude of $rv_r\sqrt{\Sigma}$ varies non-monotonically over the course of each simulation. Similarly, it appears that $rv_r\sqrt{\Sigma}$ amplitudes tend to be lower in the $\mathcal{M}=10$ simulation.

Turning to trends in the late-time behaviour as a function of both $\Omega_0$ and $\mathcal{M}$ in Figure \ref{fig:actionGrid}, it is also apparent that the amplitude of $rv_r\sqrt{\Sigma}$ varies non-monotonically with $\Omega_0$, but that generally $rv_r\sqrt{\Sigma}$ tends to become lower in amplitude as $\Omega_0$ increases. Focusing on the amplitude of $rv_r\sqrt{\Sigma}$ within the boundary layer and star, it is apparent that acoustic waves are far less significant at higher $\Omega_0$, especially for the $\Omega_0=0.8$, $\mathcal{M}=\{8,10\}$ simulations, where the amplitudes of $rv_r\sqrt{\Sigma}$ are orders of magnitude smaller than in the simulations $\Omega_0=0$ counterparts. The $\mathcal{M}=6$, $\Omega_0=0.8$ simulation is an interesting case in that it clearly displays higher-magnitude $rv_r\sqrt{\Sigma}$ within the boundary layer than within the star, while most simulations display the highest-magnitude $rv_r\sqrt{\Sigma}$ within the star, or similar magnitudes within the star and boundary layer. Similar structure can also be seen in the $\mathcal{M}=6,$ $\Omega_0=0.6$ simulation, although only throughout roughly half of the boundary layer. By comparing these plots with those of the vorticity in Figure \ref{fig:vorticityGrid}, it becomes clear that these acoustic waves are vortex-driven, and that seven vortices occupy the boundary layer in the $\Omega_0=0.6$ case, while twenty-seven vortices occupy the boundary layer in the $\Omega_0=0.8$ simulation. 

The evolution of vortensity over time in a subset of our simulations is presented in Figure \ref{fig:vortensityTimeGrid} by comparing the instantaneous vortensity to that in the initial condition. This comparison is particularly useful because vortensity is conserved along streamlines in the absence of shocks of viscosity. The influence of shocks in generating vortensity is evident in the thin filamentary structures in the the $\mathcal{M}=6,~\Omega_0=0.0$ simulation shown in the top row of Figure \ref{fig:vortensityTimeGrid}. At different times, both clockwise and counterclockwise vortices can be identified, although eventually vortices with positive vorticity dominate. Typically, simulations display multiple smaller vortices at early times, but at later times smaller numbers of more extended vortices typically remain, as expected from standard vortex dynamics where vortices of like sign merge and grow larger. We observe both compact isolated vortices, such as those in the $\Omega_0=0.8,~\mathcal{M}=6$ simulation, and `rolls', more extended in azimuth such as those shown in the $\Omega_0=0,~\mathcal{M}=10$ simulation. Such vortensity structures may or may not transform over time. For example, in the $\Omega_0=0.4,~\mathcal{M}=6$ case shown in Figure \ref{fig:vortensityTimeGrid}, many isolated vortices are visible at early times, while at later times they dissipate, leaving behind extended rolls. However, in the $\Omega_0=0.8,~\mathcal{M}=6$ simulation, isolated vortices survive until late times, and in the $\Omega_0=0,~\mathcal{M}=10$ simulation, smaller rolls populate the boundary layer at earlier times after which they seem to disappear, only for larger rolls to populate the boundary layer at later times.

Unlike \citet{2021arXiv210312119C}, we do not identify any single large vortices which drive single-armed spiral waves throughout the disk. This may be related to our outer boundary, which was at $r_{\rm out}=3r_*$, offering some support to their suggestion that larger outer boundaries of $r_{\rm out}=4r_*$ are necessary to capture such phenomena. We also use a moving mesh and significantly higher resolution in the $\phi-$direction compared to \citet{2021arXiv210312119C}, which may inhibit the production of vortensity by numerical dissipation. However, we do identify what appear to be Kelvin-Helmholtz rolls within both the star and boundary layer in the $\Omega_0=0,~\mathcal{M}=6$ simulation, as well as within the star in the $\Omega_0=0,~\mathcal{M}=8$ simulation. The vortices within the star are only visible when viewing the vorticity due to the very high density within the star.

\begin{figure}
\includegraphics[width=\columnwidth]{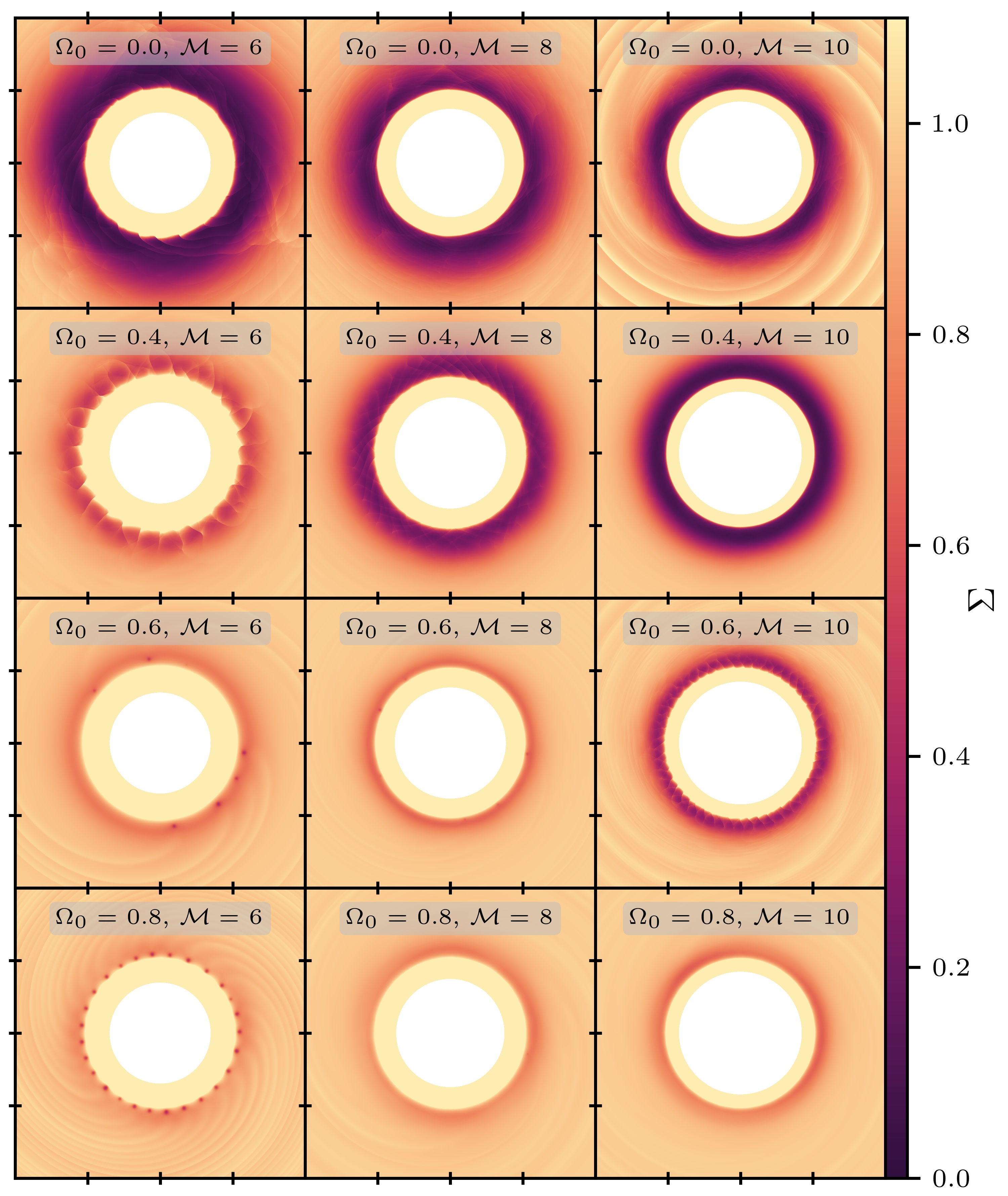}
\caption{Snapshots of surface density at $t=600\pi$ for each simulation. The columns are grouped by Mach number ($\mathcal{M}=\{6,8,10\}$), while the rows are grouped by stellar rotation rate ($\Omega_0=\{0,0.4,0.6,0.8\}$). The boundary layer becomes less depleted at high $\Omega_0$, and narrower at high-$\mathcal{M}.$}
\label{fig:surfaceGrid}
\end{figure}
\begin{figure}
\includegraphics[width=\columnwidth]{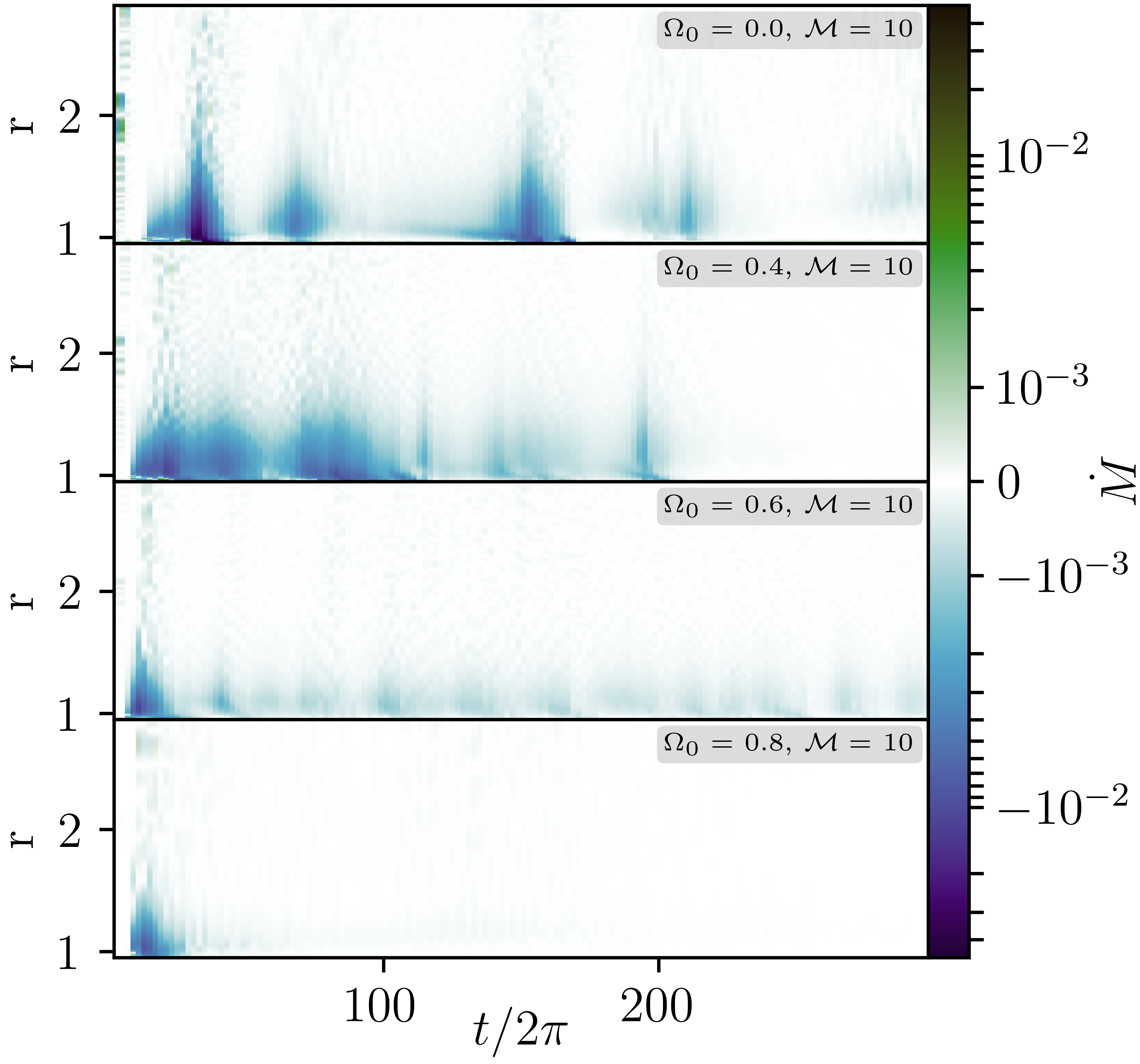}
\caption{Space-time plots of the accretion rate through the disk and boundary layer in our $\mathcal{M}=10$ simulations. We have defined $\dot{M}$, such that negative (blue) colours indicate inward flow of mass. Less accretion occurs and accretion becomes less variable for higher values of $\Omega_0$. }
\label{fig:accretion}
\end{figure}

\begin{figure}
\includegraphics[width=\columnwidth]{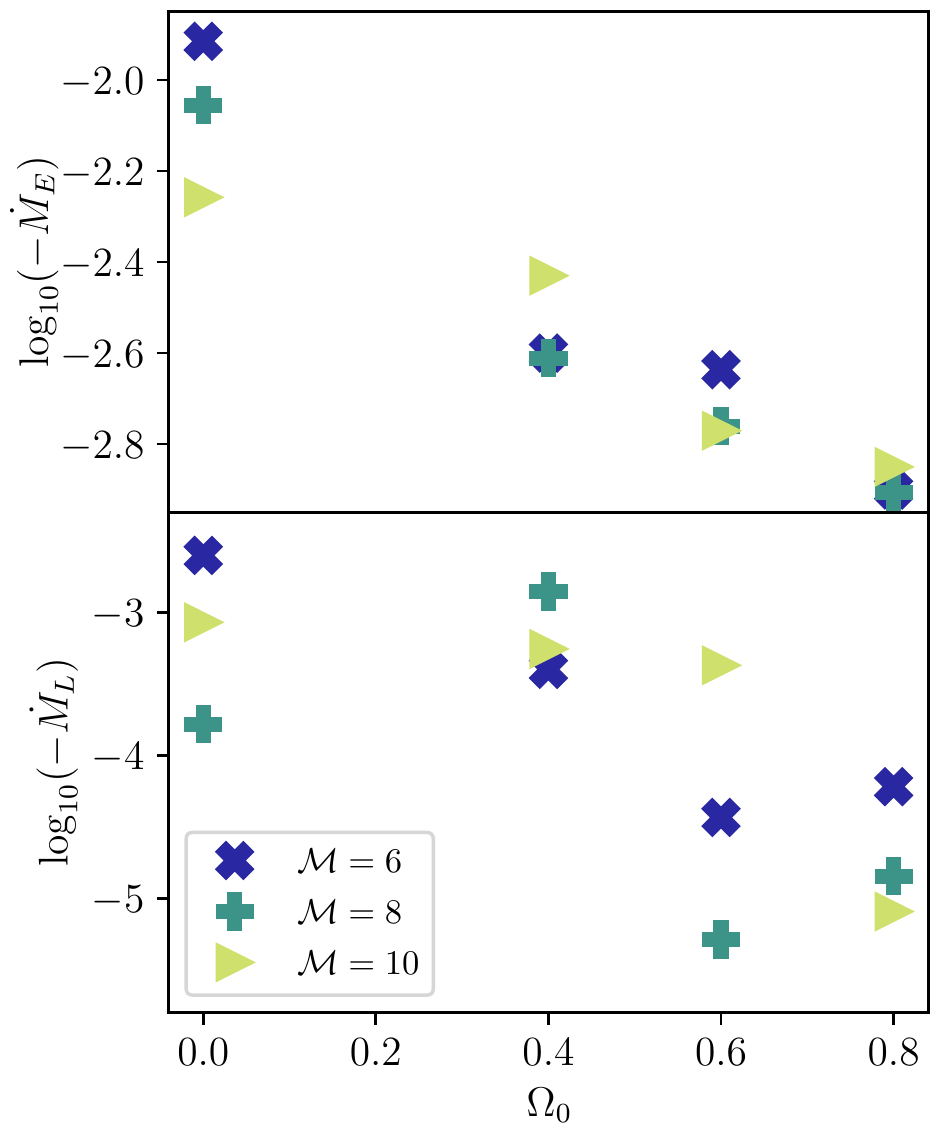}
\caption{The accretion rate through the boundary layer (at $r=r_*$) at early times (top panel) and late times (bottom panel) as a function of rotation rate. Results for $\mathcal{M}=\{6,8,10\}$ simulations are shown with blue `x' symbols, green triangles, and yellow-green `+' symbols respectively. Accretion is largely independent of Mach number, and is systematically suppressed at high $\Omega_0$, although with significant scatter in the late-time case. }
\label{fig:mdotScatter}
\end{figure}
\begin{figure*}
\includegraphics[width=\linewidth]{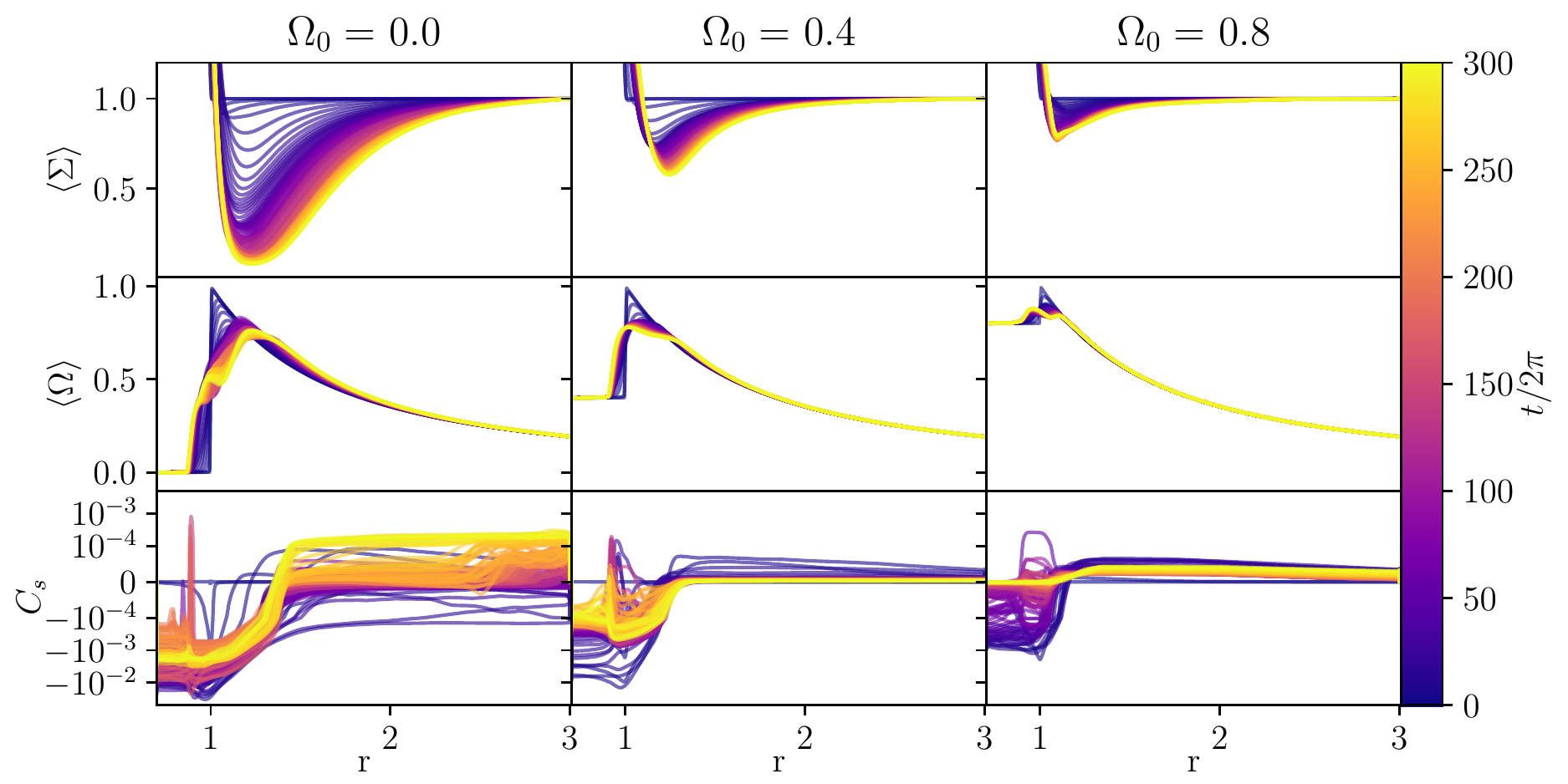}
\caption{The evolution of the time- and azimuthally-averaged surface density, angular velocity, and angular momentum current over time in three $\mathcal{M}=6$ simulations at $\Omega_0=0.0$ (left column), $\Omega_0=0.4$ (middle column), and $\Omega_0=0.8$ (right column). Line colour indicates the time at which each measurement was made in the simulation, with darker colours indicating earlier times and lighter colours indicating later times. Time averaging is carried out over periods of $4\pi$ and sampled once per time step. }
\label{fig:spacetime6}
\end{figure*}
Having observed the different ways in which vortices and waves manifest depending on $\Omega_0$ and $\mathcal{M}$, we investigate how these relate to the surface density distribution around each star. Without considering viscous or magnetohydrodynamic effects, waves are the primary mechanism for mass and angular momentum transfer, so some degree of correlation between $rv_r\sqrt{\Sigma}$ and departures in surface density from our initial distribution are natural. Qualitative correlations are indeed apparent in a comparison between the late-time profiles of $rv_r\sqrt{\Sigma}$ in Figure \ref{fig:actionGrid} and $\Sigma$ in Figure \ref{fig:surfaceGrid}. For example, cases where the boundary layer is `deeper,' or more underdense compared to the disk, tend to correspond to higher magnitudes in $rv_r\sqrt{\Sigma}$. Additionally, in simulations where coherent spiral waves propagate through the disk, they can be seen in the maps of both $\Sigma$ and $rv_r\sqrt{\Sigma}$, such as the $\Omega_0=0,0.6,~\mathcal{M}=10$ simulations, the $\Omega_0=0.6,0.8,~\mathcal{M}=6$ simulations, and more faintly in the $\Omega_0=0.8,~\mathcal{M}=8$ simulations. As vortices form in the boundary layer, they also leave an imprint in the surface density distribution, most clearly in the $\Omega_0=0.8,~\mathcal{M}=6$ simulation where each low-density point in the boundary layer corresponds to a vortex. Similarly, the $\Omega_0=0,~\mathcal{M}=10$ simulations displays $m=5$ structures in surface density, vorticity, and $rv_r\sqrt{\Sigma}$. 

The snapshots of the surface density displayed in Figure \ref{fig:surfaceGrid} also reveal a number of trends in boundary layer structure as a function of $\Omega_0$ and $\mathcal{M}$. Across all Mach numbers, the boundary layer becomes shallower as $\Omega_0$ becomes larger. This trend alone shows qualitatively that as stars rotate more quickly, waves can drive less accretion onto their surfaces. However, the extent to which the boundary layers become more `shallow' as $\Omega_0$ increases depends on Mach number. For example, the minimum late-time surface densities in $\Omega_0=0.4,~\mathcal{M}=6$ simulation are significantly larger than those in the $\Omega_0=0.4,~\mathcal{M}=10$ simulation, while the minimum surface densities in the $\Omega_0=0$
simulations at the same Mach numbers are comparable. 

The connection between $\Omega_0$ and the accretion rate onto the star is made explicit in Figure \ref{fig:accretion}, which shows the accretion rate as a function of radius time for our $\mathcal{M}=10$ simulations, where the accretion rate is defined as 
\begin{equation}
\dot{M}(r)=r\int_0^{2\pi} v_r \Sigma d\phi
\end{equation}
such that negative values correspond to a net motion of matter inwards towards the star. We note that at very early times in the low$-\Omega_0$ simulations, ripples can be seen in the accretion rate as the models settle into numerical hydrostatic equilibrium, and that these die out before instabilities manifest macroscopically. Evidently, accretion is driven intermittently, with larger bursts typically occurring earlier in the simulations. Comparing the $\Omega_0=\{0,0.4\}$ simulations, it is not obvious whether more total accretion occurs in one or the other before $t\sim400\pi$, although at later times less accretion occurs in the $\Omega_0=0.4$ simulation, and the bursts of accretion are more intense at early times in the $\Omega_0=0$ simulation. At $\Omega_0=\{0.6,0.8\}$ most of the accretion occurs at early times, almost completely stopping in the $\Omega_0=0.8$ case but continuing much more slowly in the $\Omega_0=0.6$ simulation. 

Measurements of the accretion rates through the boundary layer, at $r=r_*$, for our full set of simulations at both early ($\dot{M}_E$) and late ($\dot{M}_L$) times, are presented in Figure \ref{fig:mdotScatter}. We define `early' times as $20\pi<t<100\pi$, and `late' times  as $500\pi<t<600\pi$, and sample the averages once per timestep. We do not include $t=0$ in our early-time calculation because it takes a few orbital periods at $r_*$ for instabilities to develop and grow, and we have verified that our results are insensitive to the precise intervals chosen. At early times there is a clear trend across all Mach numbers that accretion through the boundary layer is suppressed at high $\Omega_0$. At late times, while there is still a general trend that the accretion rate onto the star decreases as $\Omega_0$ increases, there is significant scatter, and $\dot{M}_L$ is typically not monotonic with $\Omega_0$ at a given Mach number. Overall, the accretion rate at late times is much smaller in magnitude than at early times, although this is a natural result of our intrinsically transient inviscid simulations. The results of the different accretion rates displayed in Figures \ref{fig:accretion} and \ref{fig:mdotScatter} are evident in Figure \ref{fig:surfaceGrid}, as simulations with larger-magnitude accretion rates display deeper and wider surface density deficits in their boundary layers.

\begin{figure*}
\includegraphics[width=\linewidth]{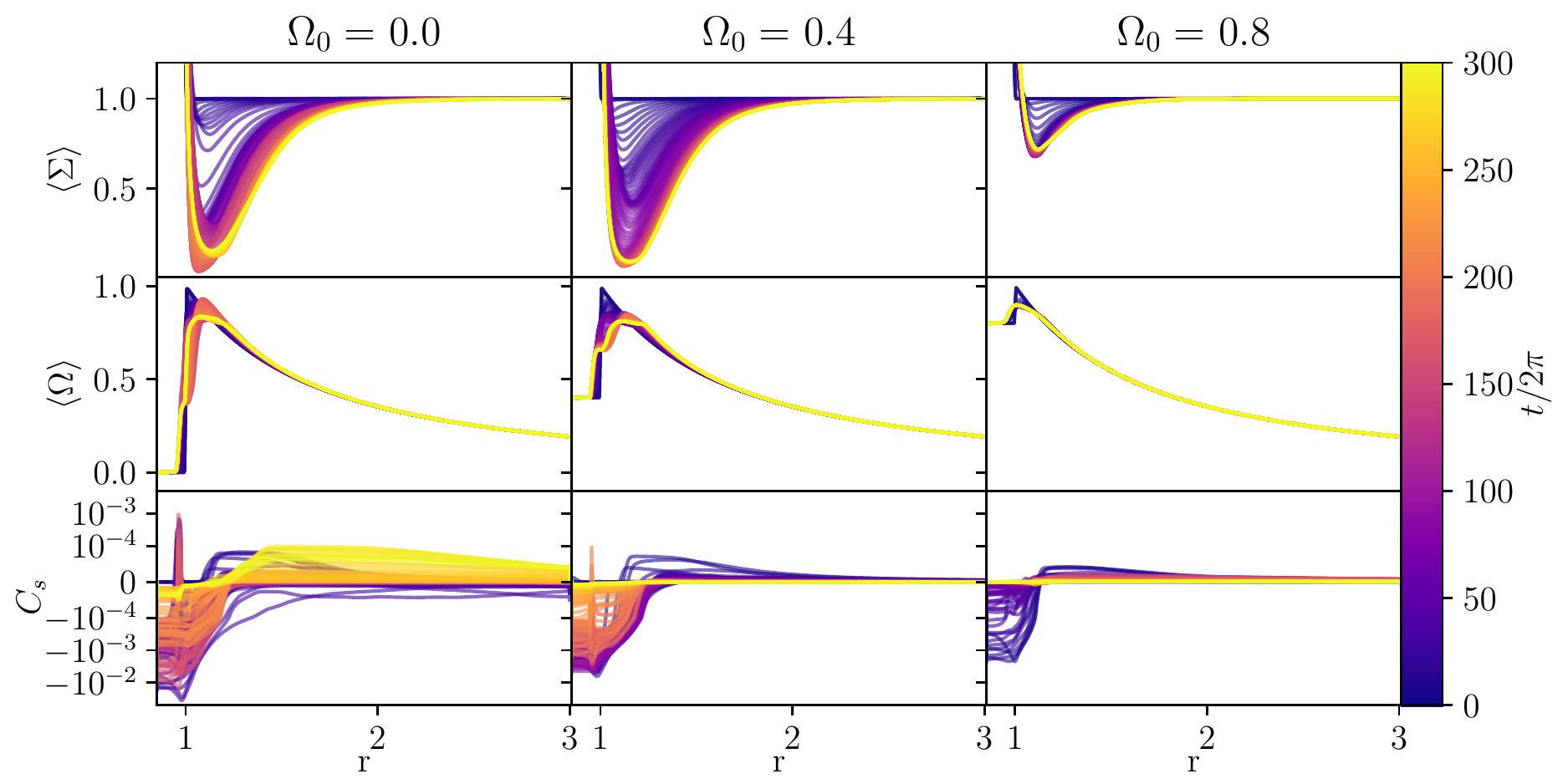}
\caption{The evolution of the time- and azimuthally-averaged surface density, angular velocity, and angular momentum current over time in three $\mathcal{M}=10$ simulations at $\Omega_0=0.0$ (left column), $\Omega_0=0.4$ (middle column), and $\Omega_0=0.8$ (right column). Line colour indicates the time at which each measurement was made in the simulation, with darker colours indicating earlier times and lighter colours indicating later times. Time averaging is carried out over periods of $4\pi$ and sampled once per time step.}
\label{fig:spacetime10}
\end{figure*}

The effects of the different accretion rates over time on the disks can be also seen, albeit indirectly,  through the azimuthally-averaged surface density distributions (for $\Omega_0=\{0,0.4,0.8\}$) shown in Figures \ref{fig:spacetime6} and \ref{fig:spacetime10} for our Mach 6 and Mach 10 simulations respectively. As each line in figures \ref{fig:spacetime6} and \ref{fig:spacetime10} is equally-spaced in time, it is clear that the surface density distribution changes more rapidly at early times, as expected from the more significant accretion rates at early times for the $\mathcal{M}=10$ simulations shown in Figure \ref{fig:accretion}. Additionally, the same general trends in both the width and depth of the surface density distribution with $\Omega_0$ and $\mathcal{M}$ can be seen in Figures \ref{fig:spacetime6} and \ref{fig:spacetime10} as in Figure \ref{fig:surfaceGrid}, but more quantitatively. We note than in the $\Omega_0=0,~\mathcal{M}=6$ simulation, depression in surface density continues to grow wider throughout the simulation, to the point where it may be affected by our fixed outer boundary condition at late times, although the depressions tend to be both shallower and narrower at higher $\Omega_0$, and narrower at higher Mach numbers, so the $\Omega_0=0,~\mathcal{M}=6$ simulation is the only case where the outer boundary has the potential to alter the surface density distribution in a meaningful way. 

The average surface densities are intrinsically linked to the average angular velocities, which follows from hydrostatic equilibrium. For example, pressure (and thus density) gradients in our initial condition balance exactly gravitational and centrifugal accelerations: while the gravitational potential of the star is held static in our initial condition, in regions where $\Omega$ decreases, $d\log{\Sigma}/dr$ must become more negative than in our initial condition, and in regions where $\Omega$ increases so must $d\log{\Sigma}/dr$. These trends are evident in Figures \ref{fig:spacetime6} and \ref{fig:spacetime10}. For example, in regions where where our initial condition had $d\Sigma/dr=0$ $(r>r_*+\delta r)$, points where $\Omega(r,t)=\Omega(r,t=0)$ correspond to minima in the associated surface density profiles. This is particularly apparent in the $\Omega_0=0,~\mathcal{M}=6$ simulation where the minimum in $\Sigma(r)$ gradually moves outward over time along with the intersection point between $\Omega(r,t)$ and $\Omega(r,0)$, and in the $\Omega_0=0.4,~\mathcal{M}=10$ simulation where similar behaviour occurs, but instead the minimum tends to move towards smaller radii. Similarly, it is clear that in each simulation $\Omega$ tends to increase in the outer accretion disk over time, corresponding the positive values of $d\Sigma/dr$ observed. We note also that a number of plateaus are visible in the azimuthally-averaged angular velocity distribution in our $\Omega_0=0,~\mathcal{M}=6$ simulation, which agree with the numerous vortices both within the star and boundary layer seen in Figures \ref{fig:vortensityTimeGrid} and \ref{fig:vorticityGrid}.

Although the distribution of angular momentum in the disk is difficult to gauge by eye because $\Omega$ tends to grow larger while $\Sigma$ becomes smaller, within the star itself both $\Omega$ and $\Sigma$ increase, suggesting that the angular momentum of the star increases over time. To explore this more quantitatively as a function of $\Omega_0$ and $\mathcal{M}$, we also plot in Figures \ref{fig:spacetime6} and \ref{fig:spacetime10} profiles of $C_s$, the angular momentum current through the domain due to waves. We recall that $C_s>0$ corresponds to angular momentum being transported outward over time, while $C_s<0$ corresponds to angular momentum being transported inwards over time. All of our simulations follow the trend that $C_s$ within the star is typically between roughly $-10^{-2}$ and $-10^{-3}$ at early times, and tends to become smaller in magnitude over time, especially for $\Omega_0>0$. The behaviour of $C_s$ in the disk over time is less universal. For example, in the $\Omega_0=0.4,~\mathcal{M}=6$ and $\Omega_0=\{0.4,0.8\},~\mathcal{M}=10$ simulations, $C_s$ tends towards values very near zero in the disk at late times, but has much larger values at early times. On the other hand, we measure larger values of $C_s$ in the disk at late times in the $\Omega_0=0,~\mathcal{M}=\{6,10\}$ simulations, and comparable values of $C_s$ at early and late times in the disk for the $\Omega_0=0.8,~\mathcal{M}=10$ simulation. In the $\Omega_0=0.8,~\mathcal{M}=6$ and $\Omega_0=0,~\mathcal{M}=10$ simulations, these larger values of $C_s$ in the disk correspond to spiral waves propagating through the disk driven by vortices in the boundary layer shown in Figures \ref{fig:actionGrid}, \ref{fig:vorticityGrid}, and \ref{fig:surfaceGrid}. However, angular momentum transport into the star in these cases is negligible, suggesting that these vortex-driven waves are not effective at spinning up the star, but instead primarily transport angular momentum out into the disk. On the other hand, clean spirals are not visible in the $\Omega_0=0,~\mathcal{M}=6$ simulation at late times, but acoustic modes are clearly active, as seen in Figure \ref{fig:actionGrid}, and able to transport angular momentum both out into the disk and into the star, even at late times.

\begin{figure}
\includegraphics[width=\columnwidth]{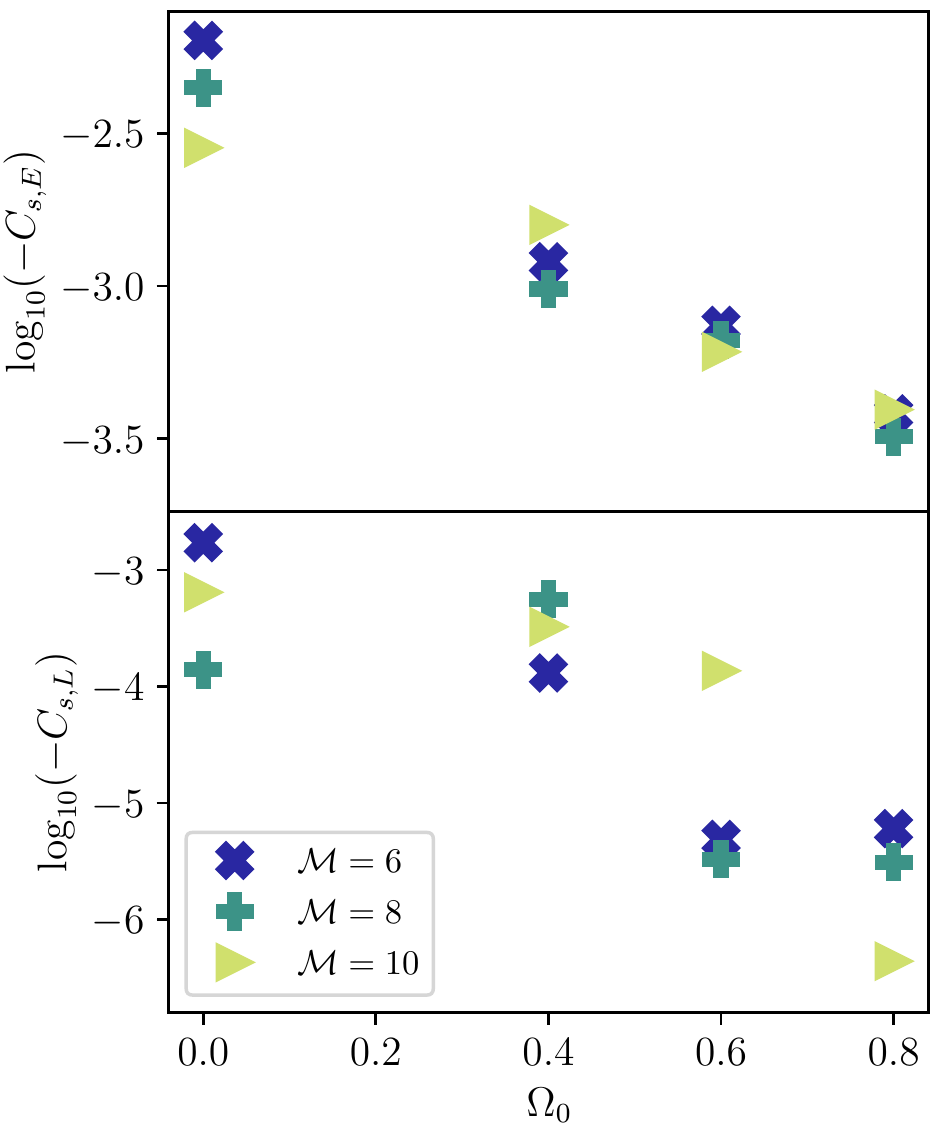}
\caption{The angular momentum current into the star due to waves at early times (top panel) and late times (bottom panel) as a function of rotation rate. Results for $\mathcal{M}=\{6,8,10\}$ simulations are shown with blue `x' symbols, green triangles, and yellow-green `+' symbols respectively. The rate at which angular momentum is transported into the star is suppressed by orders of magnitude at high $\Omega_0$, and is largely independent of Mach number.}
\label{fig:currents}
\end{figure}

To explore the extent to which the star can be spun up through acoustic interactions for different rotation rates and Mach numbers more quantitatively over our entire suite of simulations, we calculate early- and late-time averages of $C_s$ within the star ($C_{s,E}$, $C_{s,L}$), which are displayed in Figure \ref{fig:currents}. Because of our Dirichlet inner boundary condition, values of $C_s$ near the inner boundary itself are unreliable. However, as shown in Figures \ref{fig:spacetime6} and \ref{fig:spacetime10}, the profiles of $C_s$ within the star is fairly flat. Thus, we arbitrarily measure $C_s$ at the fifteenth grid cell, but have verified that errors associated with this choice are at most $\sim1\%$. We define `early' and `late' times from $20\pi<t<100\pi$ and from $500\pi<t<600\pi$ respectively, and the average was sampled once per timestep, as in Figure \ref{fig:mdotScatter}.

Clearly, spin-up is suppressed for large initial rotation rates both at early and late times. In general, spin-up is more significant at early times and is reduced steadily and monotonically with increasing rotation rate. At early times the rate of angular momentum transfer inwards decreases by roughly a magnitude between $\Omega_0=0$ and $\Omega_0=0.8$. At late times, the same overall trend of decreasing angular momentum transport at large $\Omega_0$ occurs, but it is not always monotonic (e.g. the $\mathcal{M}=8$ $\Omega_0=\{0.0,0.4\}$ cases). However, the suppression between $\Omega_0=0$ and $\Omega_0=0.8$ is much larger than for the early-time values, ranging from $\sim1.5$ to $\sim3$ orders of magnitude. In the $\mathcal{M}=10$ case, the late-time angular momentum current is not severely reduced until $\Omega_0=0.8$, while at other Mach numbers it occurs between $\Omega_0=0.4$ and $\Omega_0=0.6$. Because our simulations were inviscid and hydrodynamic, with $\dot{M}=0$ through the boundaries of the domain, it is not clear whether the early- or late-time trends are more appropriate for astrophysical scenarios, since the phenomena in our simulation are unavoidably transient. However, it is clear in either case, as stars and planets are spun up through accretion, that angular momentum transport into the star via acoustic waves becomes extremely inefficient at large rotation rates. 

\section{Discussion}\label{sec:discuss}
\subsection{Astrophysical implications}
In order to understand the implications of the results presented in Section \ref{sec:results}, especially in Figures \ref{fig:mdotScatter} and \ref{fig:currents}, we relate the quantities measured in our simulations to various astrophysical disks. One quantity of interest is the timescale for changes in the angular momentum of the accreting object, $\tau_s$. We can approximate this timescale as the ratio between the change in angular momentum required to spin-up a nonrotating object to critical rotation and the angular momentum current, or approximately 
\begin{equation}
\tau_s \approx \frac{M_*r_*^4\Omega_*}{C_s} = \frac{1}{\mathcal{C}\Sigma_0}\sqrt{\frac{M_*}{GR_*}},
\end{equation}
which is overestimated by order-unity factors due to the mass distribution of the accreting object and the neglect of the advected angular momentum current, and where $\mathcal{C}\equiv C_s/\Sigma_0R_*^4\Omega_*^2$. 

As an example, the time required to spin up Jupiter from $\Omega/\Omega_{\rm crit}=0$ to $\Omega/\Omega_{\rm crit}=0.1$ can be approximated by $\tau_s(\Omega_0=0)/10.$ For simplicity we assume present-day estimates of the mass and radius of Jupiter, a semi-major axis at formation of $5.2$ AU, and a corresponding surface density of $\sim 143~\rm{g~cm^{-2}}$ \citep{1981PThPS..70...35H}. In this case, we infer the time required to spin-up Jupiter to $\Omega/\Omega_{\rm crit}=0.1$ to be $1.4\times10^{5-7}$ years, over our full range of simulation results in Figure \ref{fig:currents}. Similarly, time required to spin up Jupiter from $\Omega/\Omega_{\rm crit}=0.6$ to $\Omega/\Omega_{\rm crit}=0.7$ ranges from $\sim2\times10^6$ years to $\sim4\times10^8$ years. As the lifetime of the solar nebula was likely less than $\sim4\times10^6$ years \citep{Wang623}, there is not enough time for Jupiter to spin up to near-critical rates, even when considering higher-density models of the Solar nebula \citep[e.g.][]{2007ApJ...671..878D} which lead to lower values of $\tau_s$. Although we do not find a strict upper limit on $\Omega/\Omega_{\rm crit}$, spinning up objects above $\Omega/\Omega_{\rm crit}\sim0.4-0.6$ takes a prohibitively long time due to inefficiencies in angular momentum transport.

In general, as $\Omega_0$ increases acoustic modes become less effective at transporting mass and angular momentum. Qualitatively, the amplitude of typical acoustic waves tends to decrease with $\Omega_0$ (see Figure \ref{fig:actionGrid}), and accretion driven by acoustic modes becomes lower in magnitude and less-variable (Figure \ref{fig:accretion}). Quasi-periodic variability has been observed in cataclysmic variable eruptions \citep[e.g.][]{2003cvs..book.....W}, which may be driven by acoustic modes generated in boundary layers \citep[e.g.][]{2012ApJ...760...22B,2021arXiv210312119C}. Additionally, acoustic waves generated in spreading layers \citep{1999AstL...25..269I,2010AstL...36..848I} on neutron star surfaces may play a key role in triggering Type I X-ray bursts in low-mass X-ray binaries \citep{2016ApJ...817...62P}. If acoustic waves are in fact an essential ingredient in these outbursts, we expect triggering these outbursts to be more difficult on faster-spinning white dwarfs and neutron stars, although simulations using a more detailed treatment of thermodynamics are necessary to make more detailed predictions. 

\subsection{Comparison with previous studies}
Unlike some previous works on the terminal spins of accreting planets, \citep[e.g.][]{2018AJ....155..178B,2020MNRAS.491L..34G}, we have considered purely hydrodynamic effects, which can operate even in the absence of magnetic fields but continue to operate in the magnetohydrodynamic case \citep[e.g.][]{2013ApJ...770...68B}. Other hydrodynamic studies of boundary layers have identified terminal rotations speeds for the central object in the sense that above certain values of $\Omega/\Omega_{\rm crit}$, ranging from $0.7$ to $0.9$, the disk transitions to a \emph{decretion} disk with $d\Omega/dr < 0$ throughout, leading to spin-down \citep{2017A&A...605A..24H,2020arXiv201206641D}. In such cases, there is a well-defined equilibrium between spin-up at low $\Omega/\Omega_{\rm crit}$ and spin-down at high $\Omega/\Omega_{\rm crit}$. However, both \citet{2017A&A...605A..24H} and \citet{2020arXiv201206641D} carried out simulations in axisymmetry, employing local shear viscosities, which cannot realistically describe angular momentum transport in boundary layers. The present study was able to capture realistic mechanisms of angular momentum transport, but does not identify a strict cut-off value of $\Omega/\Omega_{\rm crit}$, rather a steep decline in the rate at which angular momentum is transported inwards around $\Omega/\Omega_{\rm crit}\sim0.4-0.6$. The present study was limited to vertically-integrated isothermal hydrodynamics, although previous studies have found good agreement between boundary layer characteristics in two-dimensional and three-dimensional simulations \citep{2013ApJ...770...67B}.

Because our simulations were inviscid and utilised Dirichlet boundary conditions, they could not reach a true steady state between an accreting object and accretion disk, as no mass was allowed to enter or exit the domain through either boundary. This contrasts with viscous and magnetohydrodynamic simulations of boundary layers \citep{2017A&A...605A..24H,2018MNRAS.479.1528B,2020arXiv201206641D} which can couple the boundary layer to a steady accretion disk. Future work utilising either magnetohydrodynamics or viscosity which is active in the accretion disk but not in the boundary layer, as in \citet{2015A&A...579A..54H,2018MNRAS.479.1528B}, will be useful in determining the extent to which our result hold in general. 

We note that \citet{2018MNRAS.479.1528B} found that acoustic modes were insufficient to transport all of the angular momentum away from the boundary layer into the disk or star in their magnetohydrodynamic simulations as well as viscous hydrodynamic simulations with an effective $\alpha\sim0.02$ near the star. This may complicate the interpretation of our results in the stellar context.
In the case of accreting planets however, $\alpha$ may be on the order of $\sim10^{-4}$ or angular momentum transport may be altogether inviscid \citep{2017ApJ...837..163R}, so our results may be more directly applicable. 

The present study is limited by the isothermal equation of state employed, although \citep{2015A&A...579A..54H} found that global acoustic-wave-mediated mass and angular momentum transport persists when more realistic equations of state are employed. However, our isothermal equation of state precluded excitation of modes through coupling with inertial modes in the star or planet \citep{2017ApJ...835..238B}. Additionally, the entropy profile of the disk plays a key role in the development of vortices through the Rossby wave instability \citep{1999ApJ...513..805L}, which may lead to significant differences in transport due to vortex-driven modes. Additionally, a more realistic treatment of thermodynamics will be necessary to probe whether temperature changes within the star as a result of wave damping are sufficient to instigate X-ray bursts \citep{2016ApJ...817...62P}.

\section{Conclusions}\label{sec:conclusions}
We have shown that the rate at which accreting planets and stars spin up is severely reduced as those accreting objects spin more rapidly. 
The reduction in the angular momentum current into the star as $\Omega_0$ increases provides a natural and \emph{purely hydrodynamic} mechanism to limit the terminal spins of accreting stars and planets to the values measured for planets both extrasolar \citep[e.g.][]{2018NatAs...2..138B} and within the Solar System. For the case of accreting stars, particularly those in AGN disks \citep[e.g.][]{2021ApJ...910...94C,2021ApJ...914..105J,2021arXiv210212484D}, terminal spins will likely still be large enough to result in gamma-ray bursts \citep[e.g.][]{2021ApJ...906L...7P,2021ApJ...911L..19Z}.

Our study is the first to investigate the evolution via acoustic waves of boundary layers between accretion disks and rotating objects. While numerous shear-driven and vortex-driven features appear in boundary layers around rotating objects (e.g. Figures \ref{fig:actionGrid}, \ref{fig:vorticityGrid}, and \ref{fig:surfaceGrid}) as around nonrotating objects \citep[e.g.][]{2012ApJ...752..115B,2012ApJ...760...22B,2013ApJ...770...67B,2013ApJ...770...68B,2018MNRAS.479.1528B,2021arXiv210312119C}, the transport of mass and angular momentum into the accreting object is reduced for larger accretion rates as seen in Figures \ref{fig:accretion} and \ref{fig:currents}, and the resulting regions of reduced surface density are shallower and narrower for higher rotation rates as seen in Figures \ref{fig:spacetime6} and \ref{fig:spacetime10}. We have also identified a number of cases where vortex-driven modes in the boundary layer are able to effectively transport angular momentum out into the disk, but not into the accreting object. 

Unlike previous approaches to determining hydrodynamic limits on the terminal spins of planets \citep[e.g.][]{2020arXiv201206641D}, we have captured the intrinsically non-local nature of angular momentum transport in boundary layers \citep{2013ApJ...770...67B,2015A&A...579A..54H} rather than relying on a viscous \emph{local shear} stress model. Nevertheless both studies reached similar conclusions: that the rate at which stars and planets spin up through accretion decreases rapidly with the spin of the accreting object. 

\section*{Acknowledgements}
I am grateful for support from the pre-doctoral program at the Flatiron Institute Center for Computational Astrophysics, and the many stimulating conversations there which motivated this work. I thank Phil Armitage, Tad Komacek, Cole Miller, and Zeeve Rogoszinski for reading an earlier version of this manuscript and providing valuable feedback.
The simulations presented in this paper were conducted using the Rusty cluster of the Flatiron Institute and the Popeye-Simons System at the San Diego Supercomputing Center. This work was supported in part by NASA ADAP grant 80NSSC21K0649.


\bibliographystyle{mnras}
\bibliography{references.bib}




\bsp	
\label{lastpage}
\end{document}